\documentclass[10pt,journal,hyphens]{IEEEtran}

\renewcommand\IEEEkeywordsname{Index Terms}
\makeatletter
\def\ps@headings{%
	\def\@oddhead{\mbox{}\scriptsize\rightmark \hfil \thepage}%
	
	\def\@evenhead{\scriptsize\thepage \hfil \leftmark\mbox{}}%
	
	\def\@oddfoot{}%
	
	\def\@evenfoot{}}
	
\makeatother
\usepackage[hidelinks]{hyperref}
\usepackage{cite}
\usepackage[table]{xcolor}

\usepackage{float}  
\usepackage{epsfig,bm,amsmath,amssymb,graphicx,setspace}
\usepackage[numbers,sort&compress]{natbib}
\usepackage{color,placeins}
\usepackage{multirow}
\usepackage{algorithm}
\usepackage{algpseudocode}
\usepackage{colortbl}
\usepackage{color}
\usepackage{dblfloatfix}
\usepackage{turnstile}
\usepackage{multicol}
\usepackage{array}

\usepackage{enumerate}
\usepackage{turnstile}
\usepackage{subcaption}
\usepackage{arydshln,paralist}
\usepackage{soul,tabularx,booktabs}
\usepackage{csquotes}
\usepackage{multirow}
\usepackage{adjustbox}
\usepackage{float}

\usepackage{xcolor}
\usepackage{verbatim}
\usepackage[colorinlistoftodos]{todonotes} 
\usepackage{rotating}
\definecolor{usethiscolorhere}{rgb}{0.86666,0.78431,0.78431}
\usepackage{tikz}
\usetikzlibrary{mindmap}
\usetikzlibrary{calc,positioning}
\usepackage{lipsum,adjustbox}
\usetikzlibrary{decorations.pathmorphing}

\usepackage{pifont}

\makeatother

\usepackage{amsmath}
\usepackage{url}

\begin{document}

\title{FedLiTeCAN : A Federated Lightweight Transformer for Fast and Robust CAN Bus Intrusion Detection\thanks{Source code is available at \href{https://github.com/DevikaSatyan/Transformer_IDS}{https://github.com/Transformer\_IDS}}}

\author{Devika S, Pratik Narang,~\IEEEmembership{Senior Member, ~IEEE}, and Tejasvi Alladi, ~\IEEEmembership{Senior Member, ~IEEE}

\thanks{Devika S, Pratik Narang and Tejasvi Alladi are with the Department of Computer Science and Information Systems, BITS Pilani, Pilani Campus, 333031, India.  (e-mail: p20210024@pilani.bits-pilani.ac.in;pratik.narang@pilani.bits-pilani.ac.in; tejasvi.alladi@pilani.bits-pilani.ac.in).}

}
\maketitle{}

\begin{abstract}

The adoption of Intelligent Vehicles (IVs) is rising due to enhanced vehicle safety, improved navigation, and optimized performance. Intelligent Vehicles achieve efficient and smooth operations via the Controller Area Network (CAN) communication protocol; however, its lack of a built-in security system emphasizes the growing need for an Intrusion Detection System (IDS) to monitor network traffic. While prior studies have explored IDS approaches for CAN, they failed to collectively address the objectives crucial for ensuring the model's robustness, including lightweight design, fast real-time response, deployment in resource-constrained environments, unseen attack detection, and collaborative learning through Federated Learning (FL). To address this challenge, this paper introduces FedLiTECAN, a supervised intrusion detection framework that utilizes a two-layer encoder-only transformer in an FL environment. Our framework evaluation on two prominent datasets, namely the Car Hacking and Survival Analysis datasets, yielded significant performance with an overall accuracy of 98.5\%. Moreover, our framework is lightweight with a model size of 0.4MB, making it 40.56× smaller than existing baseline models. Noteably, the inference on Jetson Nano resulted in 0.608 milliseconds of detection per message, representing a speed enhancement approximately 45× faster than state-of-the-art models. A comprehensive cross-dataset analysis was performed to evaluate the framework's capability to generalize across previously unseen cyber threats, and the framework received an overall detection accuracy of 99. 996\%. These compelling results highlight FedLiTeCAN as a highly effective and practical solution for real-time intrusion detection in CAN-based vehicular networks.

\end{abstract}

\section{INTRODUCTION}
\label{sec:introduction}

Intelligent Vehicles (IVs) equipped with cutting-edge features such as real-time navigation, remote diagnostics, and Vehicle-to-Everything (V2X) \cite{devika2024vadgan} communication offer smarter transportation networks and improved traffic flow \cite{chougule2024hybridsecnet}. However, integrating sophisticated features such as real-time traffic coordination, personalized in-vehicle assistants, etc., into the vehicles increases the risks of cybersecurity threats. At the core of in-vehicle communication lies the Control Area Network (CAN) protocol that facilitates communication among various Electronic Control Units (ECUs) \cite{kang2024canival}  to exchange data efficiently. The ECU manages most of the vehicle subsystems, including airbags, acceleration, and braking mechanisms, with each electronic component having its dedicated  ECU tailored for specific tasks.

When Bosch initially introduced the CAN protocol in 1990 \cite{nguyen2023transformer}, the primary objective was to ensure reliable communication, with limited attention given to cybersecurity risks. Nevertheless, the CAN protocol does not incorporate built-in security measures such as authentication, encryption, or sender verification, significantly contributing to potential safety vulnerabilities \cite{althunayyan2024hierarchical}. However, CAN does include fundamental message integrity checks through cyclic redundancy checks \cite{chougule2024hybridsecnet}, as discussed in Section \ref{sec:background}. However, these checks are insufficient to mitigate sophisticated cyber threats. The increasing cyberattack vulnerability underscores the need for an effective Intrusion Detection System (IDS) tailored for vehicular communication environments  \cite{devika2024vadgan}. 

Modern IDSs have increasingly adopted machine learning \cite{bari2023intrusion, kidmose2024can} and deep learning techniques \cite{rai2025securing,yang2022transfer,althunayyan2024hierarchical} to build CAN intrusion detection models. Among these, Transformer-based architectures \cite{nguyen2023transformer,yang2022transfer} and other sequential models \cite{hossain2020lstm} proved effective in understanding the temporal dependencies in network data. Additionally, hybrid models that integrate Convolutional Neural Networks (CNN) with Long Short-Term Memory (LSTM) networks \cite{chougule2024hybridsecnet, nazeer2024enhancing} simultaneously capture spatial and temporal features from the network data. Complementing these strategies, anomaly-based \cite{han2018anomaly} and self-supervised learning methods \cite{jo2024intrusion,alkhatib2022can} focus on modelling the system's normal behaviour, subsequently identifying deviations that may indicate malicious activity. Few recent works have also explored Federated Learning (FL) paradigms \cite{li2020federated,zhang2025federated,zhang2023federated}, highlighting the decentralized model training and privacy-preserving solutions across distributed vehicle clients.

Despite the progress in IDS research, researchers fail to address several critical challenges. First, IDS solutions should be lightweight to ensure compatibility with resource-constrained and distributed environments \cite{zhang2025federated}, without compromising performance. Second, they should effectively handle imbalanced datasets where normal traffic vastly outnumbers anomalous traffic \cite{caivano2023marea}. Third, the IDS should demonstrate reliable performance across diverse vehicle types, including various makes, models, and operating conditions \cite{kidmose2024can}. Moreover, the IDS should timely detect previously unseen attacks, ensuring rapid response capabilities \cite{althunayyan2024robust}. However, the studies that collectively address all the aforementioned challenges remain limited. For instance, \cite{althunayyan2024hierarchical} developed a multi-stage IDS utilizing an ANN in Stage 1 and an LSTM-based autoencoder in Stage 2; however, the study did not show deployment in a resource-constrained environment, and the reported model size is comparatively larger than ours. Whereas \cite{zhang2023federated} utilized Graph Neural Networks for IDS in FL environment and demonstrated inference only on high-end GPUs. Similarly, \cite{elsayed2024boostsec} utilized the adaptive XGBoost algorithm for IDS targeting previously unseen attacks; however, the work failed to demonstrate the effectiveness of the model in a distributed setting.  
\begin{figure}
    \centering
    \includegraphics[width=0.5\textwidth]{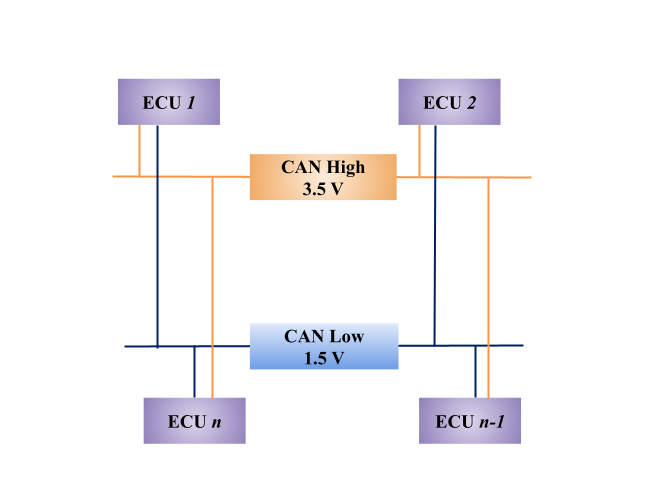}
    \caption{ECU Devices Connected through CAN Bus}
    \label{fig:canhig_low}
\end{figure}
Therefore, integration of critical factors, such as a lightweight IDS model which can be deployed to vehicle's resource-constrained embedded systems for fast response, FL to enable collaborative training while preserving privacy, and robust unseen attack detection, was an intentional design choice aimed at addressing real-world vehicular scenarios. In contrast, prior studies \cite{zhang2025federated,caivano2023marea,kidmose2024can,althunayyan2024robust,althunayyan2024hierarchical,elsayed2024boostsec,zhang2023federated} often focus on addressing only some of these aspects, leaving critical gaps in real-world applicability.

Motivated by these challenges, we developed a lightweight CAN IDS capable of detecting previously unseen attacks with high performance, fast response time, and realistic applicability in resource-constrained environments, achieved through integrating FL.

The main contributions of this work are as follows:

\begin{enumerate} \item \textbf{Lightweight Transformer-Based Intrusion Detection}: We propose FedLiTeCAN, a lightweight Transformer-based intrusion detection model for Intelligent Vehicles, featuring a compact size of 0.40 MB, approximately 40.56× smaller than existing baseline models, and achieving an overall accuracy of 98.5\% across multiple CAN datasets.

\item \textbf{Robust Generalization to Unseen Attacks}: We performed cross-dataset evaluations to assess the model’s generalizability. The model achieved a detection accuracy of 99.996\% against previously unseen attacks.

\item \textbf{Privacy-Preserving Training via Federated Learning}: The model was implemented using Federated Learning with the FedAvg strategy, where data from four vehicles were distributed across four clients to simulate a practical and privacy-preserving deployment scenario.

\item \textbf{Efficient Inference on Edge Devices}: FedLiTeCAN demonstrated fast inference capabilities, achieving a processing time of 0.608 milliseconds per sample on the Jetson Nano. This is approximately 45× faster than existing baseline models, making it well suited for deployment in resource-constrained environments.
\end{enumerate}

The remainder of the paper is organized as follows: Section \ref{sec:background} outlines the background, including CAN communication protocol, CAN Bus structure and attack taxonomy, while Section \ref{sec:related_works} reviews related work. Section \ref{sec:methodology} discusses the details of the dataset considered and the proposed FedLiteCAN architecture in centralized and federated settings. Section \ref{sec:exp} describes the experimental setup, followed by results and analysis. Finally, Section \ref{sec:conc} concludes the study.

\section{Preliminary Background}
\label{sec:background}
This section provides an overview of the preliminary background, introduces the fundamentals of the CAN communication protocol, describes the structure and operation of the CAN bus, and defines the attack taxonomy.
\subsection{CAN Communication Protocol}

In the early 1900s, communication between various vehicle components was facilitated by the Universal Asynchronous Receiver Transmitter (UART) \cite{jo2024intrusion}. However, as vehicles became more sophisticated and functional, UART could not meet the system's demands, ultimately proving inadequate. To address these limitations, Bosch developed the CAN communication protocol, which now serves as the core communication protocol in modern vehicles, particularly within the On-Board Diagnostics version two (OBD II) standard \cite{zhao2022can}. 
 
CAN is a multi-master serial communication protocol that allows multiple ECUs \cite{yang2022transfer} within a vehicle to broadcast data among themselves. Unlike the point-to-point communication protocols, CAN allows messages to be transmitted to all ECUs connected to the CAN bus \cite{chougule2024hybridsecnet}. As illustrated in Fig. \ref{fig:canhig_low}, one of the notable advantages of the CAN bus is its adaptability, allowing ECUs to be added or removed dynamically through the CAN High and CAN Low lines \cite{nguyen2023transformer}. The CAN protocol employs a priority-based arbitration mechanism using dominant bits (logical 0) and recessive bits (logical 1). This can be defined as a mechanism in which messages with more dominant bits are assigned higher priority and are transmitted first. However, adversaries can exploit this priority mechanism. For instance, an attacker could initiate a flooding attack by flooding the CAN bus with messages of higher priority, which usually contain more zeros (dominant bits) in the CAN identifier. 

\subsection{CAN Bus Format}

\begin{figure}
    \centering
    \includegraphics[width=0.5\textwidth]{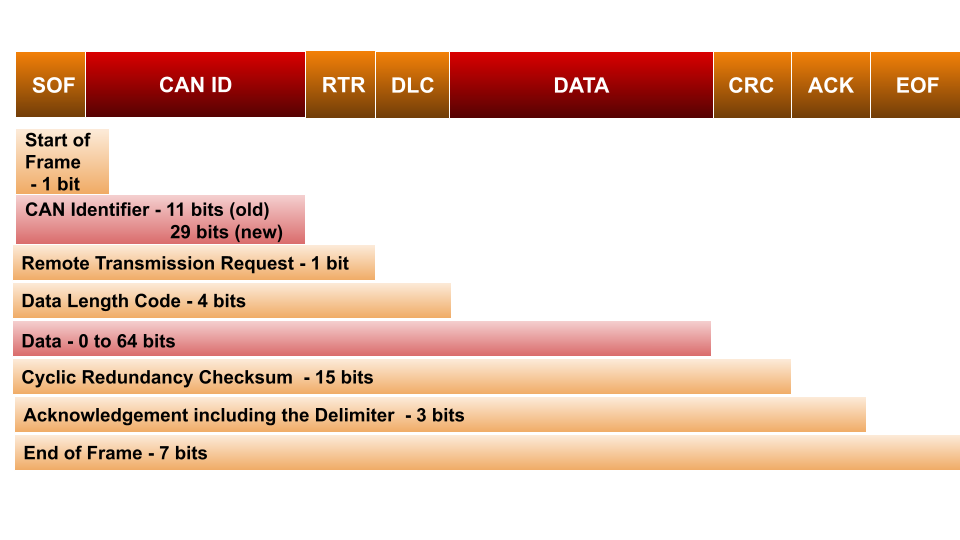}
    \caption{CAN Format}
    \label{fig:can}
\end{figure}

The structure of the CAN bus is illustrated in Fig. \ref{fig:can}. Each CAN message begins with the Start of Frame (SOF), a 1-bit field where a zero indicates the arrival of a new message to the ECU. This is followed by the CAN Identifier (CAN ID), 11 bits long in the standard format and 29 bits in the extended format. The Remote Transmission Request (RTR), a 1-bit field, allows an ECU to request data from other ECUs, whereas the Data Length Code (DLC), represented using 4 bits, specifies the length of the data field. The Cyclic Redundancy Check (CRC) field, consisting of 15 bits, is used for error checking to ensure message integrity and the Acknowledgement field (ACK), which includes 3 bits with two delimiters, indicates whether the CRC validation was successful. Finally, the message ends with the End of Frame (EOF), a 7-bit field that marks the conclusion of the transmission.
\subsection{Attack Taxonomy}
This study utilizes two distinct CAN datasets (discussed in Section \ref{sec:methodology}), each comprising various types of cyberattacks. The attack types considered are detailed as follows:
\begin{enumerate}
  
\item \textbf{Denial Of Service (DoS) / Flooding Attacks:} This refers to flooding the CAN bus with low CAN ID messages (i.e. high priority messages), preventing legitimate messages from being transmitted. 
\item \textbf{Fuzzy Attacks:} This refers to flooding the CAN bus with random CAN IDs and data payloads, unlike traditional DoS attacks.  
\item \textbf{Malfunction Attacks :} These attacks involve monitoring the CAN bus to identify valid CAN IDs and spoofing the valid CAN ID data field with the wrong data field. Two specific spoofing variations addressed in the datasets are described below.
\begin{table*}
\centering
\footnotesize
\caption{Comparison of recent automotive intrusion detection approaches.}
\renewcommand{\arraystretch}{1.2}
\begin{tabular}{|p{0.5cm}|p{1.5cm}|p{1.4cm}|p{1.5cm}|p{1.2cm}|p{1.2cm}|p{1.5cm}|p{1.5cm}|p{1.8cm}|}
\hline
\textbf{Ref.} & \textbf{Dataset Used} & \textbf{Model Used} & \textbf{Features Considered} & \textbf{Unseen Attack Detection} & \textbf{Federated Learning} & \textbf{Addressing Class Imbalance} & \textbf{Inference on Resource-Constrained Env} & \textbf{Model Size or $\#$ of Trainable Param} \\
\hline
\cite{yang2022transfer}
& Car\_hacking, CICIDS2017 dataset
& Transfer learning + CNN
& CAN ID, data[0]-[7], flag
& $\times$
& $\times$
& $\times$
& \checkmark 
& --
\\
\hline
\cite{caivano2023marea}
& Car\_hacking, Survival Analysis (only Kia), simulated dataset
& Random Forest
& data [0]-[7], flag
& $\times$
& $\times$
& \checkmark 
& $\times$
& --
\\
\hline
\cite{kidmose2024can}
& Car\_hacking, Survival Analysis, CAN-train-and-test dataset
& 16 Machine Learning Models
& timestamp, CAN ID, DLC, data[0]-[7], flag
& \checkmark 
& $\times$
& $\times$ 
& $\times$
& --
\\
\hline
\cite{han2018anomaly}
& Survival Analysis
& Statistical Detection using survival rates of CAN ID
& timestamp, CAN ID
& \checkmark 
& $\times$
& $\times$
& $\times$
& --
\\
\hline
\cite{bhavsar2024fl}
& NSL-KDD , car\_hacking
& CNN, Logistic Regression
& CAN ID, data[0]-[7], flag
& \checkmark
& \checkmark
& $\times$
& $\times$ 
& --
\\
\hline
\cite{althunayyan2024hierarchical}
& Car Hacking: Attack \& Defense
& ANN + LSTM (autoencoder)
& CAN ID, data[0]-[7]
& \checkmark 
& \checkmark
& \checkmark 
& $\times$
& 2.98 MB
\\
\hline
\cite{jo2024intrusion}
& Survival Analysis
& Unsupervised Transformer
& CAN ID, data[0]-[7]
& \checkmark 
& $\times$
& $\times$ 
& $\times$
& --
\\
\hline
\cite{alkhatib2022can}
& Car Hacking: Attack \& Defense Challenge 2020
& BERT-based bidirectional transformer 
& CAN ID, data[0]-[7], flag
& $\times$
& $\times$
& $\times$
& \checkmark 
& 2.9–3.2M params; 20–70 MB
\\
\hline
\cite{zhang2025federated}
& Car\_hacking, Ford Transit 500, IEEE Challenge 
& Two-stage Encoder-only Transformer
& CAN ID, data[0]-[7]
& \checkmark 
& \checkmark
& \checkmark
& $\times$ 
& 176,133 params (~688 KB)
\\
\hline
\cite{chougule2024hybridsecnet}
& Car\_hacking 
& Two-stage Hybrid CNN-LSTM classifier
& CAN ID, data[0]-[7], flag
& $\times$
& $\times$
& $\times$
& $\times$
& --
\\
\hline
\end{tabular}

\label{tab:ls}
\end{table*}

\item \textbf{Spoofing Attack RPM:} This refers to targeting the CAN IDs related to the engine's Revolutions Per Minute (RPM) and injecting false RPM readings to different vehicle ECUs.
\item \textbf{Spoofing Attack Gear:} This refers to targeting the CAN IDs related to the vehicle's gear shifting system. This can result in abnormal gear transitions or failures in the gear operation logic.
\end{enumerate}

\section{Related Works}

\label{sec:related_works}
The structure of our literature review is as follows: we begin by examining the IDS framework within the context of IVs using statistical techniques and popular machine learning models. Exploring deep learning approaches, including hybrid models for CAN bus intrusion, follows this. Next, we review and discuss transformer-based models demonstrating high detection performance in IDS applications. We then focus on studies employing FL frameworks for distributed intrusion detection in vehicular environments.
Next, we highlight key research gaps in recent literature, including the need to detect previously unseen attacks, handle imbalanced datasets, and efficiently deploy in resource-constrained environments. Table \ref{tab:ls} provides a detailed summary of the reviewed literature.   

Intrusion detection in CAN Bus aims to detect abnormal communication within the in-vehicle network to safeguard against potential cyber threats. Mee et al. \cite{han2018anomaly} proposed an IDS by modelling survival rates, which is the occurrence frequency of each CAN ID within defined chunks of CAN ID. This method proved effective for real-time analysis; however, the model's reliance on statistical methods was found to be less effective. While Kidmose et al. \cite{kidmose2024can} compare the performance of 16 machine learning models under different scenarios, including inclusion, exclusion, and subdivision of dataset features. However, despite working on three highly imbalanced datasets, the authors failed to address the class imbalance issue. In contrast, Danilo et al. \cite{caivano2023marea} proposed the Random Forest algorithm, demonstrating robustness to outliers and scalability to large datasets. Their work highlighted the importance of preprocessing techniques to address class imbalance issues, including padding, bitwise encoding, and under-sampling, yet it failed to demonstrate unseen attack detection.

MD et al.\cite{hossain2020lstm} introduced an LSTM-based IDS capable of binary and multi-class classification by adapting the output layer. Their study also emphasised the importance of systematic hyperparameter tuning to enhance model performance. Whereas, Amit et al. \cite{chougule2024hybridsecnet} proposed a two-stage CNN-LSTM IDS that integrates temporal and spatial features to effectively capture normal and known attack patterns. Their approach, which can be developed in environments with limited resources, increases efficiency by only enabling the second stage of the model when an attack is suspected. The model exhibited excellent performance with state-of-the-art models. In contrast, Rai et al. \cite{rai2025securing} demonstrated excellent performance across three CAN datasets using LSTM, GRU, and VGG-16 models. The authors have demonstrated generalization capabilities, binary and multi-class classification across multiple datasets, yet with no deployment in resource-constrained devices. However, \cite{hossain2020lstm} failed to consider privacy and resource-constrained deployments, whereas \cite{chougule2024hybridsecnet} was limited in unseen attack detection, distributed learning capabilities, and other gaps identified in Section \ref{sec:introduction}. 

Hyunjun et al. \cite{jo2024intrusion} proposed an IDS utilizing an unsupervised Transformer architecture. This approach enabled the model to identify previously unseen attack patterns by learning deviations from normal CAN ID sequences, and they did extensive comparisons with pre-trained Transformer and deep learning models. However, they failed to show deployment in a resource-constrained environment. Similarly, Nguyen et al. \cite{nguyen2023transformer} adopted a Transformer-based approach enhanced with transfer learning, which promised easy adaptation to new datasets. The authors also explored different input structuring strategies, analyzing the performance impact of feeding individual samples versus sequences; nevertheless, they failed to address edge deployment and privacy concerns. Whereas, Alkhatib et al. \cite{alkhatib2022can} proposed a relevant work incorporating BERT (a pre-trained Transformer) to perform IDS. The authors trained the model exclusively on normal data and achieved high detection accuracy across multiple attack types. While the bidirectional attention of BERT contributed to better performance, it also resulted in a comparatively larger model size.

Mansi et al. \cite{bhavsar2024fl} evaluated the effectiveness of CNN and Logistic Regression-based IDS models in an FL environment using the Flower framework \cite{beutel2020flower}, demonstrating results comparable with centralized training. They implemented the standard FL approach with two aggregation strategies, FedAvg \cite{mcmahan2017communication} and FedYogi \cite{reddi2020adaptive}, utilizing Jetson Xavier as a server and a Raspberry Pi as a client in two testbeds with varying numbers of clients. Whereas, Zhang et al. \cite{zhang2025federated} proposed a multi-stage IDS leveraging Transformer models within a standard FL setup. The evaluation considered both IID and non-IID (Independent and Identically Distributed) distributions (using the Dirichlet distribution) across four clients, with slightly better results obtained for IID settings. Similarly, Zhang et al. \cite{zhang2023federated} employed a multi-stage Graph Neural Network (GNN) in the standard FL framework. They have implemented FedAvg and FedProx \cite{li2020federated} strategies for server-side model aggregation under non-IID settings, with FedProx achieving the best results. In contrast, Althunayyan et al. \cite{althunayyan2024hierarchical} proposed a multi-stage IDS integrating ANN and LSTM within a hierarchical FL framework. They performed multi-level model parameter aggregation through multiple intermediate servers, improving distributed intrusion detection performance and scalability. The study explicitly focused on non-IID data distribution and was evaluated using the Flower framework. 
While \cite{bhavsar2024fl}  demonstrated effective performance for IDS, it failed to address previously unseen attack detection, and  \cite{althunayyan2024hierarchical}  limited demonstration of performance in a resource-constrained environment. In contrast,  \cite{zhang2023federated} has only reported inference on high-end GPUs with comparatively slow inference time, whereas \cite{zhang2025federated} limited its unseen attack capability and did not include deployment in a resource-constrained environment.

Despite the strong performance reported in the aforementioned CAN IDS, some critical issues still need to be addressed collectively. Notably, works such as \cite{yang2022transfer,caivano2023marea,alkhatib2022can, chougule2024hybridsecnet} did not show how well their model detected previously unseen attacks, limiting their applicability in real-world scenarios. Similarly, studies including \cite{zhang2025federated,kidmose2024can,han2018anomaly,bhavsar2024fl,alkhatib2022can,chougule2024hybridsecnet} tend to overlook the challenge of class imbalance, a common issue in real-world datasets. Among the reviewed literature, only \cite{nguyen2023transformer} has discussed inference in a resource-constrained setting. However, many existing models, such as those in \cite{alkhatib2022can}, have sizes between 20 and 70 MB, which are considerably large for deployment in such environments. These limitations underscore the need for a lightweight, fast, and accurate CAN IDS that can be deployed in real-time. 
\begin{algorithm}[ht]
\caption{FedLiTeCAN Algorithm}
\label{alg:alg}
\begin{algorithmic}[1]
\State \textbf{Input:} Total rounds $R$, clients $i = 1, \dots, N$
\State // $R$: total communication rounds, $\mathcal{D}_i$: data at client $i$, $\eta$: learning rate
\State // $X$: input batch, $H$: input projection, PE: positional encoding, p: class probabilities
\State // $\theta_{\text{CLS}}$: learnable CLS token, $e_{\text{CLS}}$: encoded CLS representation
\State // $W^Q, W^K, W^V$: projection matrices for query, key, and value; $W_1, W_2, b_1, b_2$: feed-forward neural network parameters
\State // $\alpha_y$, $\gamma$: focal loss hyperparameters, $w$: model weights
\State \textbf{Initialize:} Global model parameters $w_0$

\For{$k = 0$ to $R - 1$} \Comment{Global communication round}
    \State Server broadcasts current global weights $w_k$ to all participating clients
    \For{each client $i$ in parallel}
        \State Initialize local model with received weights: $w_i \leftarrow w_k$
        \For{each local epoch}
            \For{each mini-batch $\mathcal{B} \subset \mathcal{D}_i$}
                \State Project input: $H = \text{Linear}(X)$
                \State Concatenate CLS token: $Z \leftarrow [\theta_{\text{CLS}} \mid H] + \text{PE}$
                \For{each Transformer encoder layer}
                    \State $Q = Z \cdot W^Q$, $K = Z \cdot W^K$, $V = Z \cdot W^V$
                    \State $A = \text{softmax}\!\left(\frac{Q K^\top}{\sqrt{d_k}}\right)V$ 
                    \State $U = \text{LayerNorm}(Z + A)$
                    \State $F = \text{ReLU}(U \cdot W_1 + b_1) \cdot W_2 + b_2$
                    \State $Z = \text{LayerNorm}(U + F)$
                \EndFor
                \State$e_{\text{CLS}} \leftarrow Z[0]$
                \State Compute logits: $\hat{y} = e_{\text{CLS}} \cdot W_{\text{out}} + b_{\text{out}}$
                \State Compute $p = \text{softmax}(\hat{y})$
                \State Compute $\mathcal{L}_{\text{focal}} = \alpha_y \cdot (1 - p[y])^\gamma \cdot (-\log p[y])$
                \State Update parameters: $w_i \leftarrow w_i - \eta \nabla \mathcal{L}_{\text{focal}}$
            \EndFor
        \EndFor
        \State Client $i$ sends updated weights $w_k^i$ and metrics to the server
    \EndFor
    \State Server aggregates updates $w_{k+1}$
\EndFor
\State \textbf{Return:} Final optimized global model weights $w_R$
\end{algorithmic}
\end{algorithm}

\section{Proposed Methodology}
This section presents the datasets considered, the preprocessing steps applied, the architecture of the Transformer model implemented, and the deployment of the model within the FL framework. The algorithm described in Algorithm \ref{alg:alg} provides an overview of the methodology.
\label{sec:methodology}
    \begin{figure}
    \centering
    \includegraphics[width=0.5\textwidth]{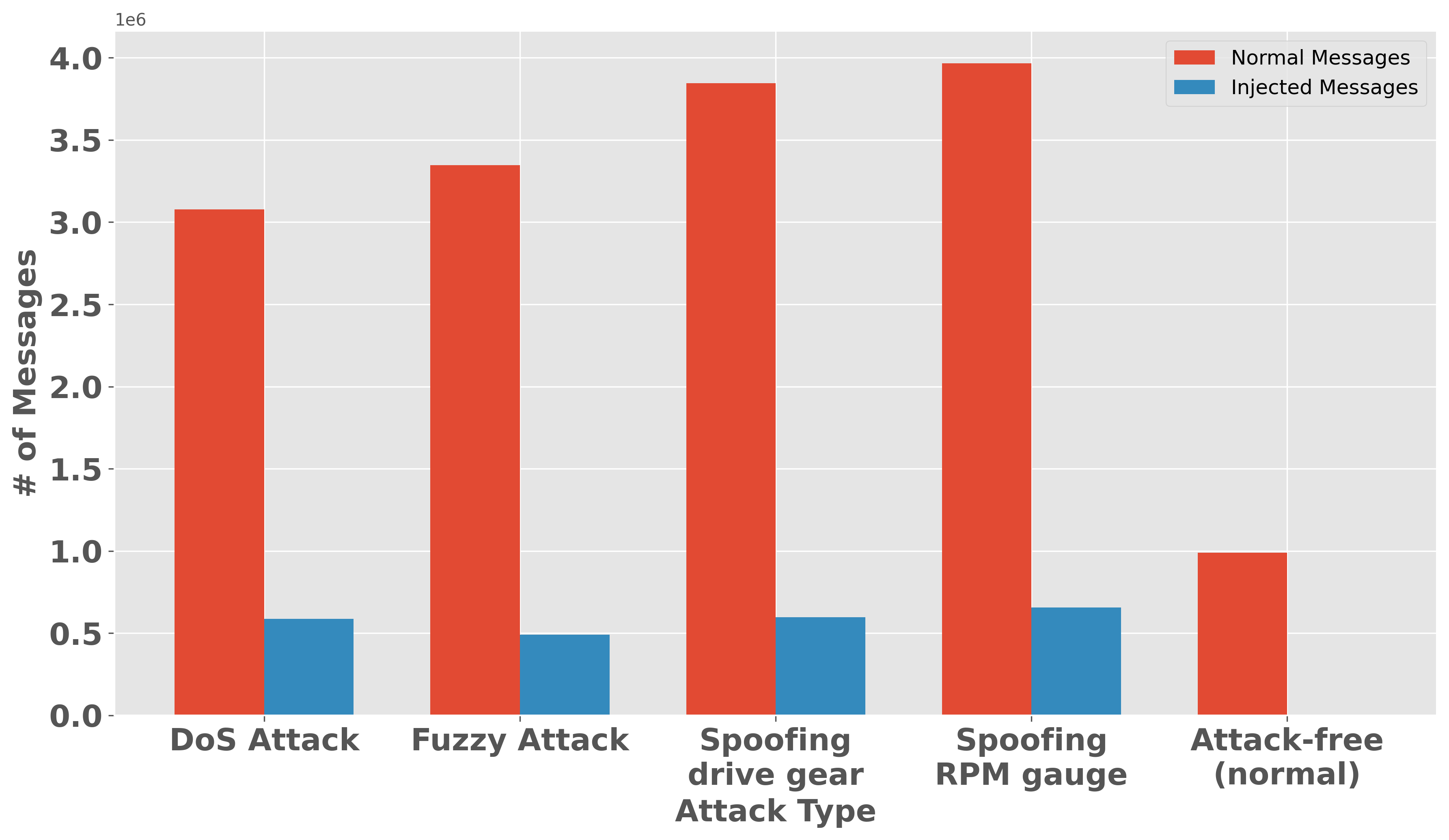}
    \caption{Overview of Car Hacking Dataset }
    \label{fig:can_overview}
\end{figure}
\subsection{Dataset Used}

The CAN protocol offers the availability of several publicly accessible datasets for intrusion detection research. In this study, we utilize two distinct datasets, each encompassing a diverse set of attack types, including various malfunction-related attacks highly relevant to real-world attack scenarios. The datasets used are as follows:

\begin{enumerate}
    \item \textbf{Car Hacking Dataset :} The Car Hacking Dataset \cite{seo2018gids} includes DoS, fuzzy, spoofing the drive gear, and the RPM gauge attacks. On average, anomalous messages were injected into the OBD-II port of real vehicular data at approximately 0.6 milliseconds. An overview of this dataset is shown in Fig. \ref{fig:can_overview}.

\begin{figure}
    \centering
    \includegraphics[width=0.5\textwidth]{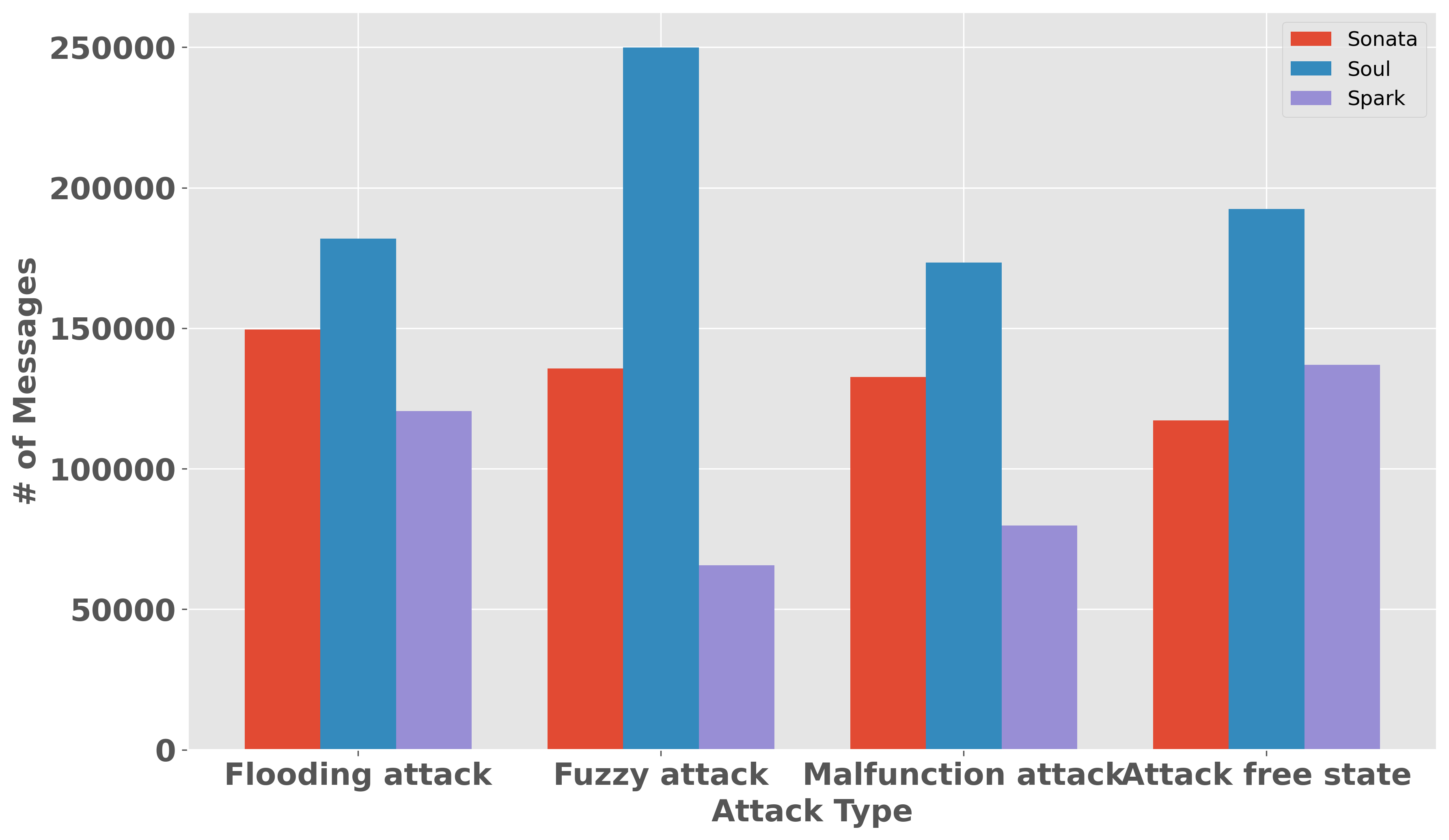}
    \caption{Overview of Survival Analysis Dataset}
    \label{fig:survival_overview}
\end{figure}

    \item \textbf{Survival Analysis Dataset for automobile IDS :} The Survival Dataset \cite{han2018anomaly} includes flooding, fuzzy, and malfunction attacks. Here, attack packets were injected for five seconds at 20-second intervals across three distinct attack types. The dataset comprises CAN traces collected from three vehicles: Sonata, Spark, and Kia. The dataset overview is depicted in Fig. \ref{fig:survival_overview}.
    \end{enumerate}

In this paper, we adopt the following naming conventions for the datasets: car\_hacking, survival\_sonata, survival\_spark and survival\_kia.

\subsection{Dataset Preprocessing}
The two datasets used in this work \cite{han2018anomaly,seo2018gids} share identical features, including a flag tag that labels each message as normal or anomalous using the letters R and T, respectively. We focused on extracting only the essential features, namely,  CAN ID, data field, and label (R or T), based on the methodologies followed in prior research. As the CAN messages are encoded initially in hexadecimal format, they were converted to decimal before being fed into the Transformer model. We adopted the approach in \cite{jo2024intrusion} to prevent overlaps, wherein each data field value was directly converted to decimal, and a constant value of 256 was added to each CAN ID post-conversion. Since not all messages are of 8-byte length, padding was applied to standardise the message length using a value of 2304 to fill incomplete fields. This value was selected to avoid conflicts with actual CAN IDs, as suggested in \cite{jo2024intrusion}.

Once preprocessed, the data were segmented into fixed-length sequences using a sliding window approach. Previous research has shown that window size can affect model performance \cite{nguyen2023transformer}, so we conducted an empirical analysis to identify an optimal window size configuration, and found that using smaller window sizes helps maintain the early information of the sequence, which is essential for successful detection.

\subsection{Transformer Architecture}
\begin{figure}
    \centering
    \includegraphics[width=0.4\textwidth]{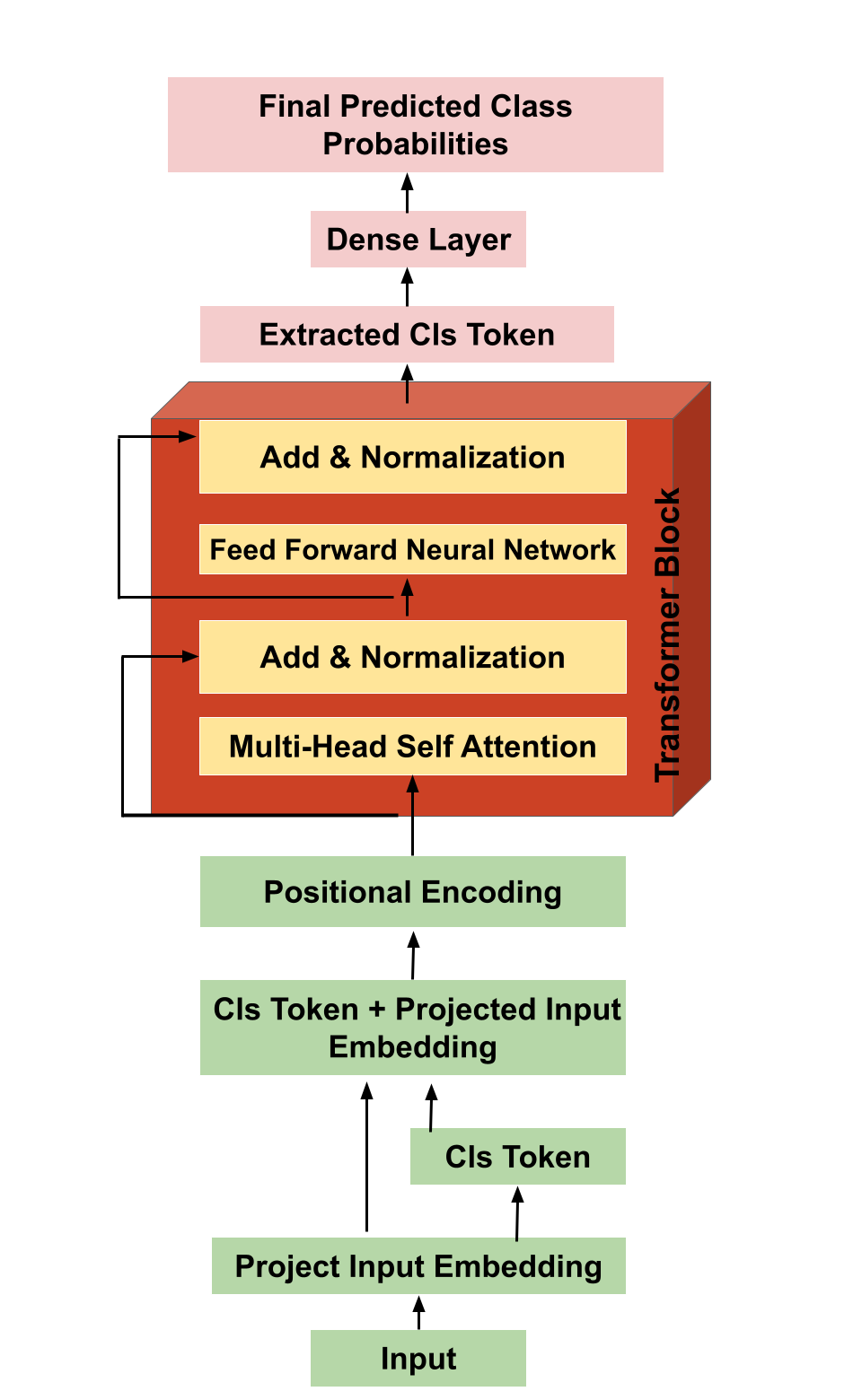}
    \caption{Proposed Encoder-only Transformer Architecture}
    \label{fig:can_trans}
\end{figure}
Fig.~\ref{fig:can_trans} illustrates FedLiTeCAN architecture for CAN intrusion detection. Let \( \mathbf{X} \in \mathbb{R}^{B \times L \times D_{\text{in}}} \) be the input sequence, where \( B \) is the batch size, \( L \) is the sequence length, and \( D_{\text{in}} \) is the input feature dimension.  First, a higher-dimensional embedding space is created by linearly projecting the input. A special learnable classification token (denoted as \( \boldsymbol{\theta}_{\text{CLS}} \)), is prepended to the sequence to serve as a summary representation for classification. Positional encodings are added to the input embeddings to encode the position of each token. The resulting sequence is passed through a stack of two transformer encoder layers. Each encoder comprises a multihead self-attention module (MHSA) and a feedforward neural network (FFNN). Residual connections and layer normalization are applied after each sub-layer to improve training stability. Within each encoder, the self-attention mechanism captures contextual relationships among all tokens, including the CLS token. The output corresponding to the CLS token is retrieved after the encoder layer and fed through a linear layer to produce class logits. These logits are then processed through a softmax layer to determine the class probabilities during inference. Focal loss \cite{lin2017focal} helps mitigate the class imbalance during training by assigning higher weights to minority samples.

\vspace{1em}

\begin{equation}
H = X \cdot W_{\text{in}} + b_{\text{in}}
\end{equation}

\begin{equation}
C = \theta_{\text{CLS}}
\end{equation}

\begin{equation}
Z = \left[ C \, \middle| \, H \right]
\end{equation}

\begin{equation}
Q = Z \cdot W^Q, \quad K = Z \cdot W^K, \quad V = Z \cdot W^V
\end{equation}

\begin{equation}
A = \text{softmax} \left( \frac{Q K^\top}{\sqrt{d_k}} \right) V
\end{equation}
\begin{figure*}
    \centering
    \includegraphics[width=0.9\textwidth]{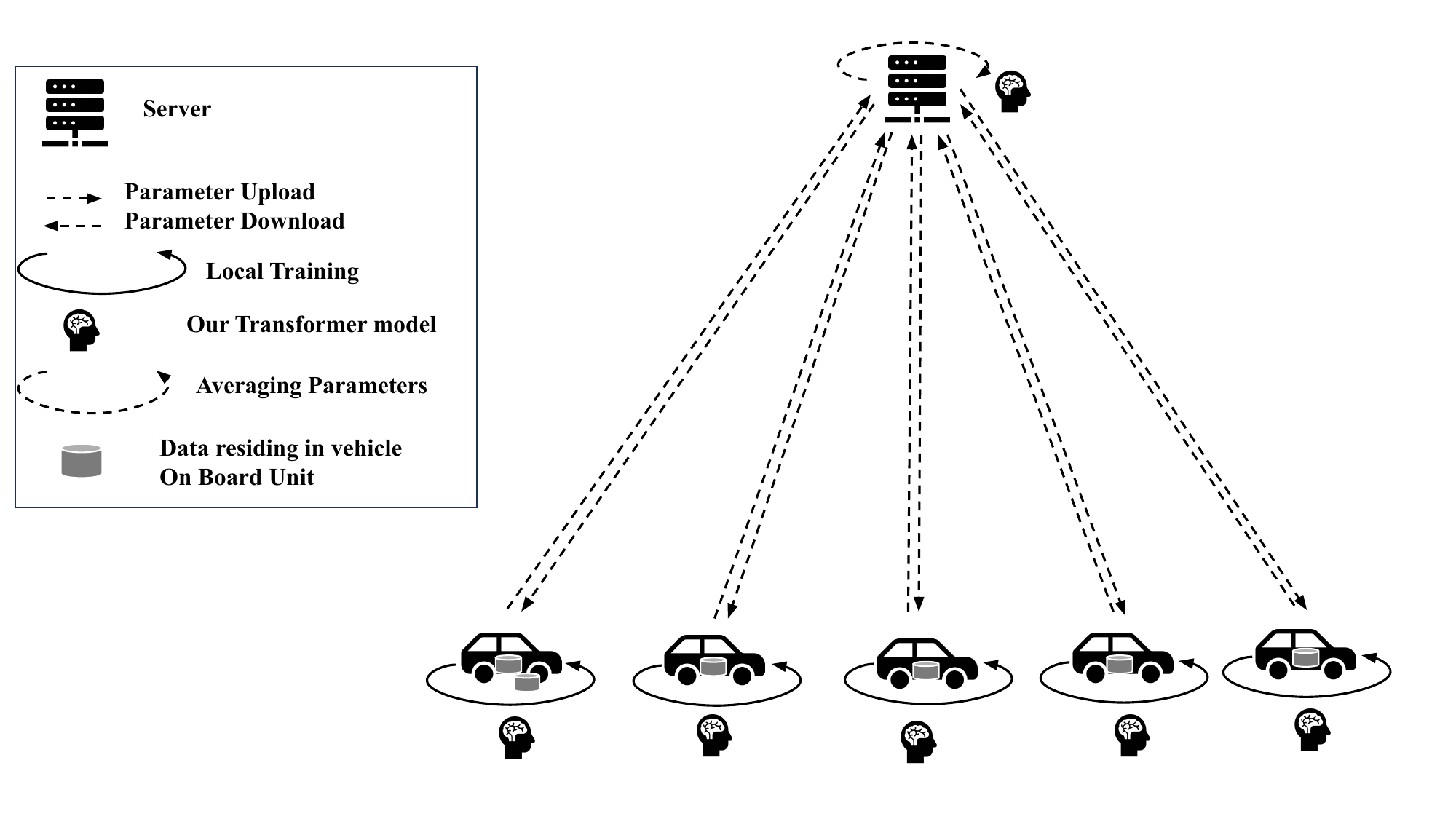}
    \caption{Proposed Transformer in FL Environment}
    \label{fig:can_transs}
\end{figure*}

\begin{equation}
U = \text{LayerNorm}(Z + A)
\end{equation}

\begin{equation}
F = \left( \max(0, U \cdot W_1 + b_1) \right) \cdot W_2 + b_2
\end{equation}

\begin{equation}
E = \text{LayerNorm}(U + F)
\end{equation}

\begin{equation}
e_{\text{CLS}} = E[0]
\end{equation}

\begin{equation}
\hat{y} = e_{\text{CLS}} \cdot W_{\text{out}} + b_{\text{out}}
\end{equation}

\begin{equation}
p = \text{softmax}(\hat{y})
\end{equation}

\begin{equation}
p_y = p[y] = \text{softmax}(\hat{y})[y]
\end{equation}

\begin{equation}
\mathcal{L}_{\text{focal}}(y, \hat{y}) = \alpha_y \cdot (1 - p_y)^{\gamma} \cdot (-\log p_y)
\end{equation}

\vspace{1em}

\textbf{Terminologies:}
\begin{itemize}
    \item $X$: Input tensor (batch size × sequence length × input dimension).
    \item $H$: Projected input embeddings.
    \item $\theta_{\text{CLS}}$: Learnable embedding vector prepended as [CLS].
    \item $C$: The CLS embedding.
    \item $Z$: Concatenation of $C$ and $H$.
    \item $W^Q, W^K, W^V$: Projection matrices for query, key, value.
    \item $d_k$: Dimension of key vectors (scaling factor).

    \item $A$: Self-attention output.
    \item $U$: Output after residual connection and normalization.
    \item $W_1, W_2, b_1, b_2$: Feedforward network parameters.
    \item $F$: Feedforward network output.
    \item $E$: Final encoder output.
    \item $e_{\text{CLS}}$: Encoded representation of the CLS token.
    \item $W_{\text{out}}, b_{\text{out}}$: Dense layer parameters.
    \item $\hat{y}$: Logits.
    \item $p$: Predicted class probabilities.
    \item $\alpha_y$: Class weight in focal loss.
    \item $\gamma$: Focal loss focusing parameter.
    \item $p_y$: Probability assigned to the true class $y$.
\end{itemize}

 The per-layer complexity of our model corresponds to $0(n^2.d + n.d^2)$, where $n$ is the sequence length and $d$ is the projection dimension. 
 
\subsection{Transformer in FL Environment}

We implemented FL using the Flower framework \cite{beutel2020flower}, where the central server coordinates training across multiple clients, each with its own local dataset. The process begins with the server initializing a global model \( w_0 \), which is sent to all participating clients. Each client then performs local training on its private data for several epochs, resulting in updated model weights \( w_k^i \), where \( i \) is the client index and \( k \) denotes the communication round. These updated weights are sent back to the central server. The server aggregates the received weights using techniques such as Federated Averaging (FedAvg) \cite{mcmahan2017communication} or FedProx (to deal with heterogeneity in federated networks) \cite{li2020federated} to compute a new global model \( w_k \). This updated global model is again distributed to the clients, and the process is repeated over multiple communication rounds until convergence. The overal setup is illustrated in Fig. \ref{fig:can_transs}.

\begin{equation}
w_k^i = \text{LocalUpdate}(w_k, \mathcal{D}_i)
\end{equation}

\begin{equation}
w \leftarrow w - \eta \nabla \mathcal{L}_i(w; \mathcal{B})
\end{equation}

\begin{equation}
\min_{w} \left[ \mathcal{L}_i(w) + \frac{\mu}{2} \| w - w_k \|^2 \right]
\end{equation}

\begin{equation}
w \leftarrow w - \eta \left( \nabla \mathcal{L}_i(w; \mathcal{B}) + \mu (w - w_k) \right)
\end{equation}

\begin{equation}
w_{k+1} = \sum_{i=1}^{N} \frac{n_i}{n} w_k^i
\end{equation}

\noindent
\textbf{Terminologies:}

\begin{itemize}
    \item \( w_k \): Global model at federated round \( k \)
    \item \( w_k^i \): Locally trained model on client \( i \) during round \( k \)
    \item \( w \): model updates
    \item \( \eta \): Learning rate used in local updates
    \item \( \mathcal{L}_i(w) \): Loss function on client \( i \)
    \item \( \mathcal{D}_i \): Local dataset on client \( i \)
    \item \( \mathcal{B} \subset \mathcal{D}_i \): Mini-batch sampled during local training
    \item \( \mu \): FedProx proximal term
    \item \( n_i \): Number of data samples on client \( i \)
    \item \( n = \sum_{i=1}^{N} n_i \): Total number of samples across all clients
\end{itemize}
The weights assigned to each client are determined by  the fraction of its data relative to the total data.

\section{Experiments}
\label{sec:exp}
This section provides details of the system specification, model hyperparameters used in this study, and a comparative analysis with the State-of-the-Art (SOTA). 
\subsection{Experimental Setup}
All experiments were carried out on an NVIDIA RTX A6000 GPU within a software environment comprising Python (version 3.10.12), PyTorch (version 2.6.0), CUDA (version 11.8), and the Visual Studio Code (VSCode) IDE. FL experiments were implemented using the Flower framework (version 1.14.0) \cite{beutel2020flower}. The access performance in resource-constrained environments, FedLiTeCAN, was also deployed on a Jetson Nano device running JetPack 4.6.6.

\subsection{Evaluation Metrics}

The evaluation metrics chosen for this work are accuracy, precision, recall, and F1-score. These metrics collectively provide a comprehensive view of the classification model's performance, addressing the correctness of predictions and the balance between identifying positives and avoiding false alarms.

\begin{equation}
\text{Accuracy} = \frac{\text{TP} + \text{TN}}{\text{TP} + \text{TN} + \text{FP} + \text{FN}}
\end{equation}

\begin{equation}
\text{Precision} = \frac{\text{TP}}{\text{TP} + \text{FP}}
\end{equation}

\renewcommand{\arraystretch}{1.2}
\begin{table}[h]
\centering
\caption{Model and Training Hyperparameters}
\scriptsize
\begin{tabular}{|l|l|p{1.2cm}|}
\hline
\textbf{Category} & \textbf{Hyperparameter} & \textbf{Value} \\
\hline
\multirow{5}{*}{General Training}
& Batch Size & 128 \\
& Learning Rate & 0.001 \\
& Epochs & 200 \\
& Early Stopping Patience & 15 \\
& Optimizer & AdamW  \\
\hline
\multirow{5}{*}{Transformer Model}
& Projection Layer Dimension & 64 \\
& Number of Attention Heads & 2 \\
& Number of Encoder Layers & 2 \\
& Dropout Rate & 0.15 \\
\hline
\multirow{2}{*}{Sequence Processing}
& Window Size & 10 \\
& Stride & 1 \\
\hline
\multirow{2}{*}{Focal Loss}
& Gamma (focusing parameter) & 2.0 \\
& Alpha (class weights) & square-root inverse class freq.  \\
\hline
\multirow{3}{*}{FL Setup}
& Clients & 4 \\
& Communication Rounds & 40 \\
& Epochs & 5\\
\hline
\end{tabular}
\label{tab:hyperparam}
\end{table}
\begin{equation}
\text{Recall} = \frac{\text{TP}}{\text{TP} + \text{FN}}
\end{equation}

\begin{equation}
\text{F1-score} = \frac{2 \times \text{Precision} \times \text{Recall}}{\text{Precision} + \text{Recall}}
\end{equation}

\renewcommand{\arraystretch}{1.2}
\begin{table}[h]
\centering
\scriptsize
\caption{Performance Metrics per Dataset and Attack Class.}

\begin{tabular}{|l|c|c|c|c|l|}
\hline
\textbf{Dataset Used} & \textbf{Acc} & \textbf{Pre} & \textbf{Recall} & \textbf{F1-score} & \textbf{Attack} \\
\hline
\multirow{5}{*}{Car\_hacking}
& 0.9991 & 0.97 & 1 & 0.99 & Normal \\
& 0.9992 & 0.99 & 1 & 1 & DoS \\
& 0.9296 & 1 & 0.93 & 0.96 & Fuzzy \\
& 0.9992 & 0.99 & 1 & 1 & Gear \\
& 0.9990 & 0.99 & 1 & 0.99 & RPM \\
\hline
\multirow{3}{*}{Survival\_sonata}
&1	&0.99	&1	&0.99	&Normal \\
& 0.9998 & 1 & 1 & 1 & Flooding \\
& 0.9873 & 1 & 0.99 & 0.99 & Fuzzy \\
& 1 & 1 & 1 & 1 & MalF \\
\hline
\multirow{4}{*}{Survival\_spark}
& 0.9851 & 0.97 & 0.99 & 0.98 & Normal \\
& 0.9881 & 0.99 & 0.99 & 0.99 & Flood \\
& 0.9081 & 0.92 & 0.91 & 0.91 & Fuzzy \\
& 0.9938 & 0.99 & 0.99 & 0.99 & MalF \\
\hline
\multirow{4}{*}{Survival\_kia}
& 0.9991 & 0.99 & 1 & 0.99 & Normal \\
& 1 & 1 & 1 & 1 & Flood \\
& 0.9681 & 1 & 0.97 & 0.98 & Fuzzy \\
& 0.9961 & 0.90 & 1 & 0.95 & MalF \\
\hline
\end{tabular}
\label{tab:centralized_results}
\end{table}
\subsection{Hyperparameters}
\begin{table*}[htbp]
\centering
\scriptsize
\caption{Comparison of Performance Metrics Across SOTA for CAR HACKING dataset.}
\renewcommand{\arraystretch}{1.2}
\begin{tabular}{|p{1.0cm}|p{2.22cm}|p{1.1cm}|p{1.1cm}|p{1.1cm}|p{1.1cm}|p{1.1cm}|}
\hline
\textbf{Ref.} & \textbf{Model} & \textbf{Accuracy} & \textbf{Precision} & \textbf{Recall} & \textbf{F1-score} & \textbf{Attacks} \\

 \hline
 \multirow{4}{*}{\cite{seo2018gids}} 
 &  & 97.9 & 96.8 & 99.6 & - & DoS \\
 &  & 98.0 & 97.3 & 99.5 & - & Fuzzy \\
 &  GAN& 98.0 & 98.3 & 90.0 & - & gear \\
 &  & 96.2 & 98.1 & 96.5 & - & rpm \\
\hline
\multirow{1}{*}{\cite{chougule2024hybridsecnet}}

& CNN-LSTM & 0.9957 & 0.9958 & 0.9957 & 0.9958 & Overall \\
\hline
\multirow{1}{*}{\cite{zhang2025federated}}

&  & 1 & 1 & 1 & 1 & DoS
\\
& &0.994712 & 0.993583 & 0.994147 & 0.997552 & Fuzzy \\
& Transformer & 1 & 1 & 1 & 1 & gear \\
&  & 0.999906 & 1 & 0.999953 & 0.999974 & rpm \\
\hline
\multirow{1}{*}{\cite{ashraf2020novel}}

& LSTM (autoencoder) &0.99& 0.99 & 1.00& 0.99 & Overall\\

\hline

\multirow{4}{*}{\cite{rai2025securing}}
&  & 1& 1& 1 &1&DoS\\
&VGG-16 & 0.99 &0.94&0.99 &0.90 & Fuzzy\\
&  & 0.96& 1& 0.96 &0.98&gear\\
&  & 1& 1& 1 &1 & rpm\\
\hline

\multirow{4}{*}{\cite{xin2025luft}}
&  & 0.999
& -& - &0.997&DoS\\
& FFT-CNN & 0.976 & -&- &0.981 & Fuzzy\\
&  & 0.992 & - & - &0.989&gear\\
&  & 0.994 & -&- &0.991 & rpm\\
\hline

\multirow{4}{*}{\cite{cheng2023desc}}
&  & - & 96.69& 95.62 &96.08 &DoS\\
& TCN (autoencoder) & - & 95.25& 95.33 &95.27 & Fuzzy\\
&  & - & 97.88 &97.84 &97.86&gear\\
&  & - & 96.41 &96.00 &96.17 & rpm\\
\hline
\multirow{4}{*}{\cite{wang2023effective}}
&    &0.9980 &0.9953 &0.9721 &0.9836 &DoS\\
& Graph Convolutional Network  & 0.9970& 0.7804& 0.8392 &0.8088 & Fuzzy\\
&  &0.9980 &0.9996& 0.8920 &0.9427&gear\\
&  & 0.9980 &0.9932& 0.9355& 0.9635 & rpm\\
\hline
\multirow{4}{*}{\cite{shi2024ids}}
&  & 99.48 & 99.98 & 98.28 & 99.12 & DoS \\
& LSTM-CNN &99.51 &99.99 &99.23 &99.61 & Fuzzy \\
&  &99.57 & 99.91 & 99.10 & 99.50 & gear \\
&  & 99.26& 99.98& 98.38& 99.17 & rpm \\

\hline
\multirow{5}{*}{\cite{nguyen2023transformer}} 
&  & - & 0.99932 & 0.99996 & 0.99964 & Normal \\
&  & - & 1 & 1 & 1 & DoS \\
& Transformer & - & 0.99991 & 0.99979 & 0.99985 & Fuzzy \\
&  & - & 0.99549 & 0.9645 & 0.99597 & gear \\
&  & - & 0.99838 & 0.99707 & 0.99772 & rpm \\
\hline
\multirow{5}{*}{\textbf{Ours}} 
&  & \textbf{0.9991} & \textbf{0.97} & \textbf{1} & \textbf{0.99} & \textbf{Normal} \\
&  & \textbf{0.9992} & \textbf{0.99}& \textbf{1} & \textbf{1} & \textbf{DoS} \\
& \textbf{Transformer} & \textbf{0.9926} & \textbf{0.99} & \textbf{0.93} & \textbf{0.96} & \textbf{Fuzzy} \\
&  & \textbf{0.9992}& \textbf{0.99} & \textbf{1} & \textbf{1} & \textbf{gear} \\
&  & \textbf{0.999} &\textbf{ 0.99 }&\textbf{ 1} & \textbf{0.99} &\textbf{rpm} \\
\hline
\end{tabular}

\label{tab:performance_comparison}
\end{table*}

\begin{figure}
    \centering
    \includegraphics[width=0.3\textwidth]{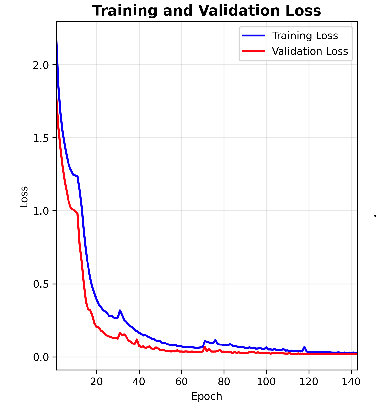}
    \caption{Graph showing that the model is not overfitting }
    \label{fig:overfit}
\end{figure}
The hyperparameters used in our experiments are listed in Table \ref{tab:hyperparam}. These values were finalized after conducting an extensive ablation study. We experimented with various configurations, including different numbers of encoder layers [2,4,8], attention heads [2,4,8], and projection layer dimension [64, 128]. While large encoder layers, attention heads, and projection dimensions increased performance, it also increased model size to 8,04,229 trainable parameters, which is heavy for a CAN IDS. Notably, we observed that a lightweight configuration with a 2-layer encoder, two attention heads, and a projection dimension of 64 achieved comparable performance, drastically reducing the parameter count to 103,716 (approximately 0.104M).

We observed an early stopping at epoch 143 during a 200-epoch training process, using a patience value of 15, as illustrated in Fig. \ref{fig:overfit}.
In the FL setup, we selected five local epochs and 40 communication rounds to approximate the 200 epochs trained in the centralized approach closely. We also explored configurations such as (20 rounds, 10 epochs) and (200 epochs, 1 round), but we got the best results with 40 rounds and five epochs.

\subsection{Comparison of Centralized Results with SOTA}

This section compares the performance of FedLiTeCAN with SOTA models on the Car\_hacking and Survival datasets. The results obtained by FedLiTeCAN are illustrated in Table \ref{tab:centralized_results}.

\subsubsection{Car\_Hacking Dataset}
To evaluate the performance of FedLiTeCAN on the car\_hacking dataset, we have compared the results with SOTA models from \cite{seo2018gids,chougule2024hybridsecnet,zhang2025federated,nguyen2023transformer,ashraf2020novel,rai2025securing,xin2025luft,cheng2023desc,wang2023effective,elsayed2024boostsec}, as illustrated in Table \ref{tab:performance_comparison}.  The study in  \cite{seo2018gids} and \cite{chougule2024hybridsecnet}  exhibited strong performance across all attack types; 
however, they failed to address class imbalance and detect previously unseen attacks. Whereas \cite{ashraf2020novel} and \cite{rai2025securing} demonstrated high detection accuracy but could not illustrate the model's lightweight nature and models' deployment in resource-constrained environments. The study in \cite{xin2025luft} integrated Fast Fourier Transform (FFT) with CNN, but failed to report precision and recall, with no distributed learning to illustrate the collaborative nature of vehicles. On the other hand, the study in \cite{wang2023effective} illustrated poor performance for fuzzy attacks with a precision of 78.04\% and an F1-score of 80.88\%. In contrast, \cite{shi2024ids} illustrated an excellent detection rate across all the attack types but at a higher latency. Similarly, \cite{zhang2025federated} and \cite{nguyen2023transformer} utilized transformer-based models and reported strong results with overall performance; however, their substantial model size hinders deployment in resource-constrained environments. In contrast, our FedLiTeCAN demonstrates competitive performance and is suitable for real-world deployment.

\begin{table}[h]
\scriptsize
    \centering
     \caption{Overall results of Survival Analysis Dataset with baseline SOTA}
    \begin{tabular}{|c|c|c|c|c|c|}
    \hline
    \textbf{Ref} & \textbf{Model} & \textbf{Acc} & \textbf{Pre} & \textbf{Recall} & \textbf{F1score} \\ 
    \hline
    \cite{rai2025securing} & VGG-16 & 0.9425 & 0.8825 & 0.9425 & 0.892 \\ 
    \hline
    Ours & Transformer & 0.9854 & 0.9792 & 0.9867 & 0.9829 \\ 
    \hline
    \end{tabular}
   
    \label{tab:overallsurvival}
\end{table}

\begin{table}[ht]
\centering
\caption{Comparison of performance with SOTA on Survival Sonata Dataset.}
\scriptsize
\renewcommand{\arraystretch}{1.2}
\begin{tabular}{|l|l|c|c|c|c|l|}
\hline
\textbf{Ref} & \textbf{Model} & \textbf{Acc} & \textbf{Pre} & \textbf{Recall} & \textbf{F1-score} & \textbf{Attack} \\
\hline
\multirow{3}{*}{\cite{han2018anomaly}} 
& Statistical & - & - & 0.9861 & - & Flood \\
&  & - & - & 0.9735 & - & Fuzzy \\
&  & - & - & 0.9995 & - & MalF \\
\hline
\multirow{13}{*}{\cite{jo2024intrusion}} 
& Uni-GPT & - & 0.409 & 1 & 0.580 & Flood \\
&  & - & 0.32029 & 0.996 & 0.484 & Fuzzy \\
&  & - & 0.27711 & 0.818 & 0.414 & MalF \\
\cline{2-7}
& Bi-GPT & - & 0.365 & 1 & 0.535 & Flood \\
&  & - & 0.25200 & 0.996 & 0.402 & Fuzzy \\
&  & - & 0.23800 & 0.985 & 0.383 & MalF \\
\cline{2-7}
& CAN-BERT & - & 0.401 & 1 & 0.572 & Flood \\
&  & - & 0.300 & 0.996 & 0.460 & Fuzzy \\
&  & - & 0.283 & 1 & 0.414 & MalF \\
\cline{2-7}
& CAN-Former & - & 0.178 & 1 & 0.302 & Flood \\
&  & - & 0.103 & 1 & 0.187 & Fuzzy \\
&  & - & 0.094 & 1 & 0.172 & MalF \\
\cline{2-7}
& Transformer & - & 0.987 & 1 & 0.993 & Flood \\
&  & - & 0.736 & 0.993 & 0.846 & Fuzzy \\
&  & - & 0.805 & 0.962 & 0.877 & MalF \\
\hline
\multirow{3}{*}{\cite{zhang2025federated}} 
& \multirow{3}{*}{Transformer} & 1 & 1 & 1 & 1 & Flood \\
&  & 0.998 &0.910 & 0.952 &0.977 & fuzzy \\
&  & 1 & 1 & 1 & 1 & MalF \\
\hline
\multirow{4}{*}{\textbf{Our}} 
&  & \textbf{1 }& \textbf{0.99} & \textbf{1} & \textbf{0.99} & \textbf{Normal} \\
& \textbf{Transformer} & \textbf{0.99998 }&\textbf{ 1} & \textbf{1 }&\textbf{ 1} &\textbf{Flood}\\
&  & \textbf{0.9873} & \textbf{1 }& \textbf{0.99} & \textbf{0.99} & \textbf{Fuzzy} \\
&  & \textbf{0.9992} & \textbf{0.99 } & \textbf{1 }&\textbf{ 1 }& \textbf{MalF} \\
\hline
\end{tabular}
\label{tab:sonata_comp}
\end{table}

\subsubsection{Survival\_Sonata Dataset}
Table \ref{tab:overallsurvival} illustrates the overall results on the survival analysis dataset with baseline SOTA model reported in \cite{rai2025securing}. The performance comparison in Table \ref{tab:sonata_comp} highlights the superiority of FedLiteCAN over existing SOTA methods in \cite{han2018anomaly}, \cite{jo2024intrusion}, and \cite{zhang2025federated}. In \cite{jo2024intrusion}, the authors have fine-tuned a few pre-trained transformers such as Unidirectional GPT \cite{radford2018improving}, Bidirectional GPT \cite{nam2021intrusion}, CANBERT \cite{alkhatib2022can} and CAN-Former IDS \cite{cobilean2023anomaly} which exhibited significantly lower precision, recall, and F1-scores (often below 0.6 and in some cases as low as 0.10–0.30). Notably, even the Transformer model in \cite{jo2024intrusion} showed better results, but has not addressed the class imbalance problem. The high-performing Transformer in \cite{zhang2025federated} achieves an accuracy of 0.9988 and an F1-score of 0.9774, yet they have not recorded the results of how the model detected normal data. In \cite{han2018anomaly}, the model detected only recall results and omitted performance measures for normal data. This demonstrates that our model, FedLiTeCAN, delivers highly reliable detection of multiple attacks, especially flood attacks.
 
\begin{table}[ht]
\centering
\caption{Comparison of performance with SOTA on Survival Spark Dataset.}
\scriptsize
\renewcommand{\arraystretch}{1.2}
\begin{tabular}{|l|l|c|c|c|c|l|}
\hline
\textbf{Ref} & \textbf{Model} & \textbf{Acc} & \textbf{Pre} & \textbf{Recall} & \textbf{F1score} & \textbf{Attacks} \\
\hline
\multirow{4}{*}{\cite{han2018anomaly}} 
& \multirow{4}{*}{Statistical} & - & - & 0.9235 & - & Flood \\
&  & - & - & 0.8708 & - & Fuzzy \\
&  & - & - & 0.9983 & - & MalF \\
\hline
\multirow{3}{*}{\cite{zhang2025federated}} 
& \multirow{3}{*}{Transformer} & 1 & 1 & 1 & 1 & Flood \\
&  & 0.957 & 0.947 & 0.736 & 0.828 & Fuzzy \\
&  & 1 & 1 & 1 & 1 & MalF \\
\hline
\multirow{4}{*}{{\textbf{Ours}}} 
& \multirow{4}{*}{\textbf{Transformer} }& \textbf{0.9851 }& \textbf{0.97} & \textbf{0.99} & \textbf{0.98} & \textbf{Normal} \\
&  & \textbf{0.9881} & \textbf{0.99} & \textbf{0.99} & \textbf{0.99} & \textbf{Flood} \\
&  & \textbf{0.9081} & \textbf{0.92} & \textbf{0.91} & \textbf{0.91} & \textbf{Fuzzy} \\
&  & \textbf{0.9938} & \textbf{0.99} & \textbf{0.99} & \textbf{0.99} & \textbf{MalF} \\
\hline
\end{tabular}
\label{tab:spark_comp}
\end{table}

\subsubsection{Survival\_Spark Dataset}

The comparison in Table \ref{tab:spark_comp} shows that FedLiTeCAN is much better, achieving 11.76\% higher recall and 5.83\% higher F1-score than the Transformer in [10]. Furthermore, our model delivers a more balanced and complete evaluation compared to the statistical approach in [1], which lacks key performance metrics such as accuracy, precision and recall, and records comparatively lower recall (average 0.9308). 
\begin{table}[ht]
\centering
\scriptsize
\caption{Comparison of Performance with SOTA on Survival Kia Dataset.}
\renewcommand{\arraystretch}{1.2}
\begin{tabular}{|l|l|c|c|c|c|l|}
\hline
\textbf{Ref} & \textbf{Model} & \textbf{Acc} & \textbf{Pre} & \textbf{Recall} & \textbf{F1score} & \textbf{Attacks} \\
\hline
\multirow{3}{*}{\cite{han2018anomaly}} 
& \multirow{3}{*}{Statistical} & - & - & 0.997& - & Flood \\
&  & - & - & 0.989 & - & Fuzzy \\
&  & - & - & 0.999 & - & MalF\\
\hline
\multirow{4}{*}{\cite{caivano2023marea}} 
& \multirow{4}{*}{Random Forest} &  & 0.9999 & 1 & 0.9999 & Normal \\
&    && 1 & 1 & 1 & Flood \\
& & 0.9997 & 1 &  0.9956 & 0.9978 & Fuzzy \\
&  & & 0.9994 &  1 & 0.9979 &  MalF \\
\hline
\multirow{3}{*}{\cite{zhang2025federated}} 
& \multirow{3}{*}{Transformer} &  1 & 1 & 1 & 1 & Flood \\
&  & 0.974 & 0.990 & 0.956 & 0.973 & Fuzzy \\
&  &  0.977 & 0.978 & 0.978 & 0.993 & MalF \\
\hline   
\multirow{4}{*}{{\textbf{Ours}}} 
& \multirow{4}{*}{\textbf{Transformer}} & \textbf{0.999} & \textbf{0.99} & \textbf{0.99} & \textbf{0.99} & \textbf{Normal} \\
&  & \textbf{1} & \textbf{1} & \textbf{1} & \textbf{1} & \textbf{Flood} \\
&  & \textbf{0.9681} & \textbf{0.97} & \textbf{0.98} & \textbf{0.98} & \textbf{Fuzzy} \\
&  & \textbf{0.9961} & \textbf{0.9} & \textbf{1} & \textbf{0.95} & \textbf{ MalF} \\
\hline
\end{tabular}
\label{tab:kia_comp}
\end{table}
\subsubsection{Survival\_Kia Dataset}

The results in Table \ref{tab:kia_comp} demonstrate how FedLiTeCAN outperformed SOTA baseline methods in performance and robustness. The statistical method proposed in \cite{han2018anomaly} provides only recall values for individual attacks and lacks overall performance metrics. 
Although the model in \cite{caivano2023marea} achieves slightly higher accuracy, it lacks analysis on overfitting and scalability, and particularly in constrained hardware environments. However, our model consistently produces strong and balanced results across all evaluation metrics. In particular, our model performs well against Normal, Flood, Fuzzy, and Malfunction attacks, achieving F1-scores of up to 0.99, accuracy of up to 0.9991, precision of up to 1, and recall of up to 1. This represents a significant improvement over the Transformer model in \cite{zhang2025federated}, which displays relatively lower performance measures (accuracy 0.974–0.977 and F1-score 0.973), indicating limitations in its capacity to generalize across a variety of attack types.

\subsection{Evaluation of Federated Learning Results}

To illustrate the effect of implementation in an FL environment, we tried with different numbers of clients ranging from 4 to 6. Increasing the clients affected the performance, and we chose four as the best number of clients, where we distributed the four datasets - car\_hacking, survival\_sonata, survival\_spark, and survival\_kia to 4 different clients. To showcase the effect of a non-IID data scenario, the car\_hacking dataset was further partitioned among 5 and 6 clients. As summarized in Table \ref{tab:fl_comp}, the best results were achieved with four clients, where client 1 utilized 1/3rd of the dataset from the car\_hacking dataset and clients 2 to 4 used the entire survival analysis dataset. Additionally, FedAvg performed better than FedProx across different proximal term values (0.1 and 0.01) by 0.15\% increase in accuracy and required approximately 1.103 hours less training time. 
\begin{table}[h]
\centering
\scriptsize
\caption{Evaluation of Results in FL Setup}
\begin{tabular}{|p{1.3cm}|p{2.22cm}|p{1.2cm}|}
\hline
\textbf{\# of Clients} & \textbf{Averaging Strategy Used} & \textbf{Performance Drop} \\
\hline
6 & FedAvg  & 9.59\% \\
\hline
5 & FedAvg & 6.56\% \\
\hline
\multirow{3}{*}{\textbf{4}} & \textbf{FedAvg}  & \textbf{6.46\%} \\
& FedProx (0.1)  & 6.78\% \\
& FedProx (0.01)  & 7.05\% \\
\hline
\end{tabular}
\label{tab:fl_comp}
\end{table}

\subsection{Detection of previously unseen attacks}

We employed a cross-dataset evaluation technique to demonstrate the model's ability to detect previously unseen attacks. After training the model on the survival\_sonata dataset, it was evaluated using attack classes from the car\_hacking datasets. This method showed a strong ability of the model to generalize, with an average detection accuracy of 99.996\%. FedLiTeCAN exhibits a minor improvement of 0.006\% in detecting previously unseen attacks compared to \cite{althunayyan2024robust}. Furthermore, FedLiTeCAN has a distinct advantage over \cite{althunayyan2024robust}, because our use of cross-dataset validation provides more rigorous and comprehensive evidence of the model's generalization performance across different vehicular attack scenarios.

\subsection{Model Size Comparison with SOTA}
FedLiTeCAN is a highly lightweight model, comprising 0.104 million trainable parameters, a compact model size of 0.40 MB, and 95.183 MFLOPs per batch (batch size is 128), making it well-suited for resource-constrained deployment when compared with encoder-only transformer models such as Tiny-BERT \cite{jiao2019tinybert} and Mobile-BERT \cite{sun2020mobilebert}. In contrast, \cite{althunayyan2024hierarchical} has a size of 2.98 MB, while \cite{alkhatib2022can} ranges between 20 MB and 70 MB. Even the relatively compact model in \cite{zhang2025federated} has a size of 688 KB, whereas \cite{xin2025luft} comprises 0.169M parameters.
This demonstrates that our model is smaller than \cite{zhang2025federated} and significantly smaller than \cite{althunayyan2024hierarchical}  and \cite{alkhatib2022can}, making it highly suitable for deployment in environments with limited computational resources.

\subsection{Model Inference Time comparison with SOTA in Resource-Constrained and Intensive Platforms}

Table \ref{tab:inference_time} presents the inference latency of FedLiTeCAN across different hardware platforms, from resource-constrained devices (Colab CPU and Jetson Nano) to high-performance environments (GPU). FedLiTeCAN resulted in 0.018 ms in GPU, 0.076 ms in Colab CPU, and 0.608 ms in Jetson Nano for per-sample inference.
\begin{table}[h]
\centering
\scriptsize
\caption{Inference Time Comparison Across Multiple Environments.}
\begin{tabular}{|l|c|c|c|}
\hline
\textbf{Ref.} & \textbf{Inference Time (ms)} & \textbf{Environment} & \textbf{Samples/sec} \\
\hline
\cite{xin2025luft} & 967 & GPU & $>$10,000 \\
\hline
\cite{nguyen2023transformer} & 11.6 & CPU & $>$10,000 \\
\hline
\cite{alkhatib2022can} & 0.8 -- 3.46 & GPU & -- \\
\hline
\cite{zhang2025federated} & 4.43 & GPU & 7,218 \\
\hline
\multirow{3}{*}{\textbf{Ours}} 
& \textbf{0.018} & \textbf{GPU} & \textbf{5,54,670} \\
& \textbf{0.076} & \textbf{Colab CPU} & \textbf{1,31,680} \\
& \textbf{0.608} & \textbf{Jetson Nano} & \textbf{16,440} \\
\hline
\end{tabular}
\label{tab:inference_time}
\end{table}
 It is important to note that inference highly depends on the hardware configuration and whether latency is measured per sample, per sequence, or batch. Our proposed model demonstrates a substantial advantage over SOTA models in processing throughput. On GPU, it processes 5,54,670 samples per second, which is over 76× faster compared to \cite{zhang2025federated}, which processes only 7,218 samples per second. Even on a Colab CPU, our model delivers 1,31,680 samples per second, over 13× faster than \cite{nguyen2023transformer}. Our inference in Jetson Nano maintains a competitive 16,440 samples processed per second, still outperforming \cite{zhang2025federated}, \cite{nguyen2023transformer}, and \cite{xin2025luft}, despite those studies utilizing CPU or GPU-based environments. Our model achieves up to 44× faster processing on GPU when compared to \cite{alkhatib2022can}, where both times are recorded per sample. This confirms FedLiTeCAN's efficiency and real-time capability in high-performance and edge environments.

\section{Conclusion}
\label{sec:conc}
This research presents a practical, lightweight intrusion detection system for the in-vehicle Controller Area Network (CAN) bus. The proposed model employs a compact two-layer encoder-only Transformer architecture with a model size of 0.40MB, approximately 40.56× smaller than state-of-the-art models. To ensure real-world applicability, the model is deployed within a Federated Learning (FL) framework, demonstrating its suitability for collaborative learning. The proposed model effectively detects a broad range of cyberattacks from multiple CAN datasets, with an average accuracy of 98.5\%. Furthermore, the IDS maintains an inference latency of 0.608 milliseconds, making it 45× faster in messages processed per second than existing baseline approaches. Cross-dataset evaluation further confirms the model's generalization ability, achieving 99.996\% accuracy against previously unseen attacks.

Future work will focus on extending the evaluation to additional CAN datasets, generating synthetic intrusion scenarios, and investigating advanced FL strategies such as differential privacy and homomorphic encryption to enhance the security and robustness of the proposed system.

\bibliographystyle{IEEEtranN}
{\footnotesize
\bibliography{ref}}

\end{document}